\documentclass[aps,pre,amsmath,amssymb,reprint,longbibliography,superscriptaddress]{revtex4-1}
\usepackage{hyperref}
\usepackage{graphicx}
\usepackage{dcolumn} 
\usepackage{color}
\usepackage{bm} 

\definecolor{apsblue}{rgb}{0.18,0.19,0.57}
\definecolor{darkblue}{rgb}{0.2,0.1,0.5}
\definecolor{darkgreen}{rgb}{0.1,0.6,.1}
\definecolor{darkred}{rgb}{0.7,0.0,.1}

\begin{document}

\title{Translational and rotational dynamical 
heterogeneities in granular systems}

\author{Binquan Kou}
\affiliation{
School of Physics and Astronomy, Shanghai Jiao Tong University, 800 Dong
Chuan Road, Shanghai 200240, China}

\author{Yixin Cao}
\affiliation{
School of Physics and Astronomy, Shanghai Jiao Tong University, 800 Dong
Chuan Road, Shanghai 200240, China}

\author{Jindong Li}
\affiliation{
School of Physics and Astronomy, Shanghai Jiao Tong University, 800 Dong
Chuan Road, Shanghai 200240, China}

\author{Chengjie Xia}
\affiliation{
School of Physics and Astronomy, Shanghai Jiao Tong University, 800 Dong
Chuan Road, Shanghai 200240, China}

\author{Zhifeng Li}
\affiliation{
School of Physics and Astronomy, Shanghai Jiao Tong University, 800 Dong
Chuan Road, Shanghai 200240, China}

\author{Haipeng Dong}
\affiliation{
Department of Radiology, Ruijin Hospital, Shanghai Jiao Tong University School of
Medicine, Shanghai 200025, China}

\author{Ang Zhang}
\affiliation{
Department of Radiology, Ruijin Hospital, Shanghai Jiao Tong University School of
Medicine, Shanghai 200025, China}

\author{Jie Zhang}
\affiliation{
School of Physics and Astronomy, Shanghai Jiao Tong University, 800 Dong
Chuan Road, Shanghai 200240, China}
\affiliation{
Institute of Natural Sciences, Shanghai Jiao Tong University, Shanghai 200240,
China}

\author{Walter Kob}
\email{walter.kob@umontpellier.fr}
\affiliation{Laboratoire Charles Coulomb,
University of Montpellier and CNRS, Montpellier 34095, France}

\author{Yujie Wang}
\email{yujiewang@sjtu.edu.cn}
\affiliation{
School of Physics and Astronomy, Shanghai Jiao Tong University, 800 Dong
Chuan Road, Shanghai 200240, China}

\affiliation{
Materials Genome Initiative Center, Shanghai Jiao Tong University, 800 Dong
Chuan Road, Shanghai 200240, China}

\affiliation{
Collaborative Innovation Center of Advanced Microstructures, Nanjing University, Nanjing, 210093, China}

\begin{abstract}
We use X-ray tomography to investigate the translational and rotational
dynamical heterogeneities of a three dimensional hard ellipsoids
granular packing driven by oscillatory shear. We find that particles
which translate quickly form clusters with a size distribution
given by a power-law with an exponent that is independent of the
strain amplitude. Identical behavior is found for particles that are
translating slowly, rotating quickly, or rotating slowly. The geometrical
properties of these four different types of clusters are the same as
those of random clusters. Different cluster types are considerably
correlated/anticorrelated, indicating a significant coupling between
translational and rotational degrees of freedom. Surprisingly these
clusters are formed already at time scales that are much shorter than
the $\alpha-$relaxation time, in stark contrast to the behavior found
in glass-forming systems.

\end{abstract}

\maketitle

The relaxation dynamics of most disordered materials, such as
glass-forming liquids, polymers, foams, granular materials,
differs significantly from the Debye behavior found in
simple liquids in that it shows a marked non-exponential time
dependence~\cite{binder_11,cavagna_09}. Although several mechanisms
can give rise to this itype of time dependence, e.g. for polymeric
systems it is the chain connectivity, it is often the local
disorder of the particle arrangement that is the origin for this
behavior~\cite{ediger_00,richert_02,berthier_11}, i.e.~the fact that
each particle has a different local environment makes that the relaxation
dynamics depends strongly on the particle considered. Previous studies
have shown that the slowly (or quickly) relaxing particles are not
distributed uniformly in space but instead form clusters.  This effect,
usually named ``dynamical heterogeneity'' (DH) is nowadays believed to
be a key ingredient to understand the $\alpha-$process of glass-forming
systems, and hence the phenomenon of the glass-transition.  As a
consequence, many studies have been carried out to study the nature of
the DH, in particular how the size of the clusters depends on temperature
(or density) of the system~\cite{hurley_95,ediger_00,kegel_00,weeks_00,richert_02,berthier_11,cicerone_95,donati_98,maccarrone_10}.

Usually DH are associated with the translational degrees of freedom
(TDOF) of the particles. The particles in molecular systems and
granular materials have, however, also rotational degrees of freedom
(RDOF). Since these are coupled with the TDOF it is evident that
they will be important for the relaxation dynamics of the system
as well~\cite{zheng_14,chang_97,chong_09,mishra_15,vivek_17}. However, in
practice it is difficult to probe the RDOF in molecular systems
since experiments do not allow to track directly the
orientation of individual particles. As a consequence only
indirect experimental probing of the RDOF has been been possible so
far~\cite{cicerone_95,chang_97} and most of our current knowledge of
them comes from computer simulations~\cite{kammerer_98,qian_99,mazza_07}. The
situation is not much better for the case of granular materials
since these are usually opaque and hence it is very challenging to
probe in three dimensions (3d) the displacement and reorientation of the
particles~\cite{mueth_00,richard_03,ciamarra_07,borzsonyi_12,fu_12,ando_13,harrington_14}.
Because of the non-spherical shape of the particles and the presence
of friction, there is often a strong coupling between the TDOF and
RDOF, making the experimental study of the DH for both TDOF and
RDOF indispensible for a thorough understanding of the relaxation
dynamics~\cite{mohan_02}. In the present work we thus use X-ray tomography
to investigate these DH in a 3d granular packing driven by oscillatory
shear, making it to the best of our knowledge the first experimental
investigation to probe simultaneously both types of DH.

\begin{figure}[tbh]
{\centering
\includegraphics[width=0.48\textwidth]{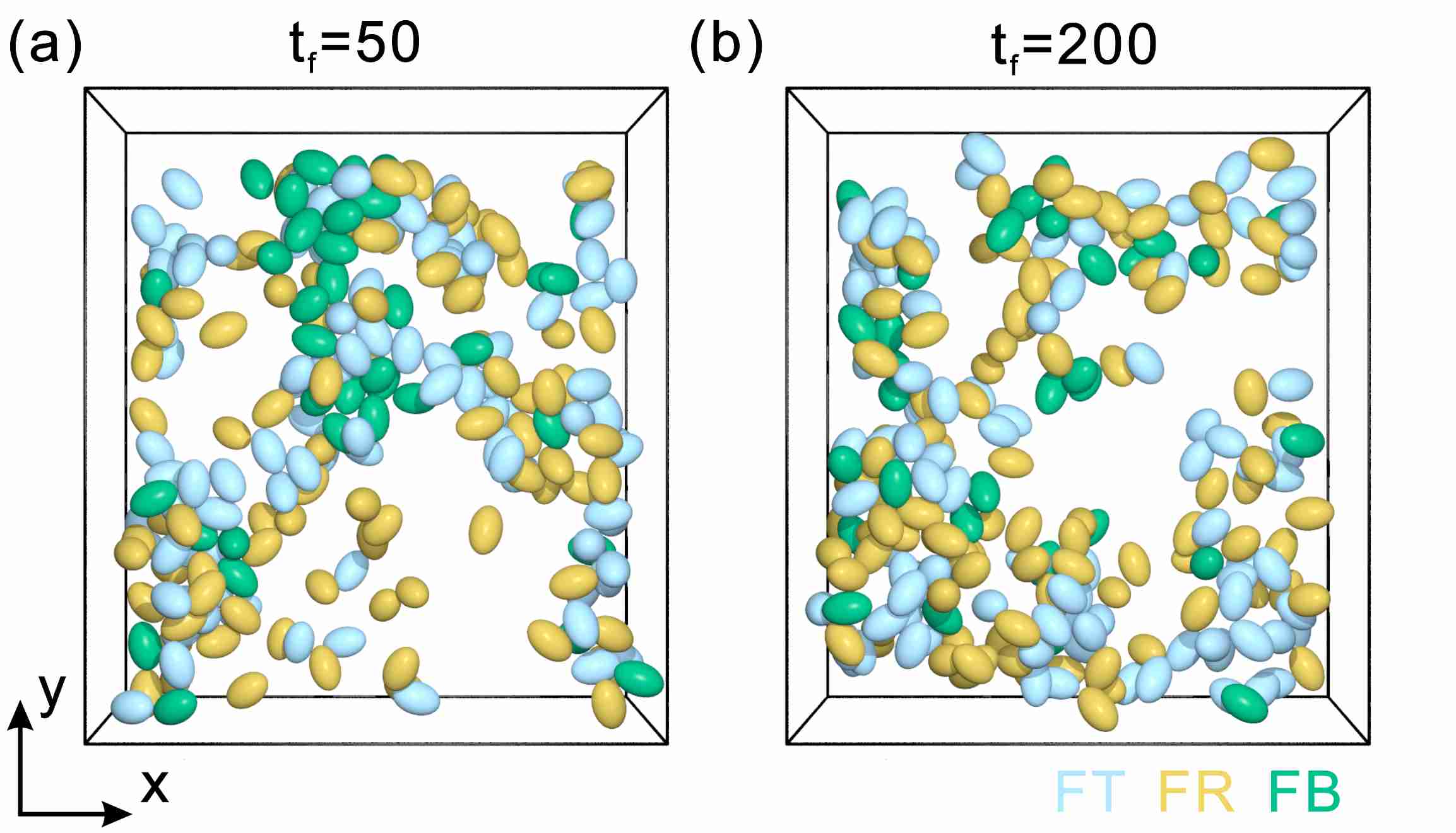}
}
\caption{
Snapshots showing the fast translating (FT, blue) and fast rotating
(FR, yellow) particles ($\gamma=0.10$(L), $t=1000$). Also shown are
the particles that are FT as well as FR (FB, green). Note that both
type of particles form clusters and that these clusters overlap
significantly. Panels (a) and (b) are for the filter time $t_f=50$
and $t_f=200$, respectively.
}
\label{fig1_FT_FR_snapshots}
\end{figure}

Our system consists of 4100 hard prolate ellipsoids made of polyvinyl
chloride with an aspect ratio of 1.5 (polydispersity 0.9\%), i.e.~a shape
that makes the crystallization of the system difficult~\cite{donev_04}
and allows for a rather strong T-R coupling. The dimension of the minor
axis of the particles is 2$b$=12.7mm and in the following we will use
$b$ as the unit of length.  The particles are in a rectangular box of
dimension $40.2b \times  43b \times 22.6b$ that can be sheared in the
$y-$direction. More details of the setup are given in~\cite{kou_17}. We
drive the system to a steady state by cycling it many times at all
strain amplitude $\gamma$ investigated ($\gamma=0.26$, 0.19, 0.10, and
0.07, see SI for details). Subsequently the position and orientation
of all particles are determined by X-ray tomography scans. Scans were
made after each complete cycle, thus giving a stroboscopic view of the
dynamics with the time unit of one shear cycle, and in the SI we show
the mean squared displacement for the TDOF and the RDOF which allows to
get an idea on the relavant time and length scales in the system.

To probe the DH we have tracked the particles for a ``filter time''
$t_f$ and determined the distribution of their T-and R-displacements
(see SI). Fast (slow) translation particles are defined as the 10\%
fastest (slowest) particles in this distribution and we denote these
particles as FT and ST. The same was done for the RDOF, thus allowing to
define the fast (slow) rotating particles, FR and SR. We have verified
that the results presented below do not depend in a significant manner
on these definitions.

In Fig.~\ref{fig1_FT_FR_snapshots} we show typical snapshots of the FT
and FR particles (blue and yellow, respectively) for two values of the
filter time $t_f$. One recognizes that both populations form clusters
which shows that the T and R dynamics are spatially very heterogeneous.
A significant part of the FT particles are also FR (FB, marked in green),
indicating that the translational and rotational motion are significantly
coupled. Similar conclusions are reached for the slowly moving particles,
i.e.~the ST and SR (see SI). In the following we will make a quantitative
characterization of these DHs.

\begin{figure}[tbh]
\centering
\includegraphics[width=0.5\textwidth]{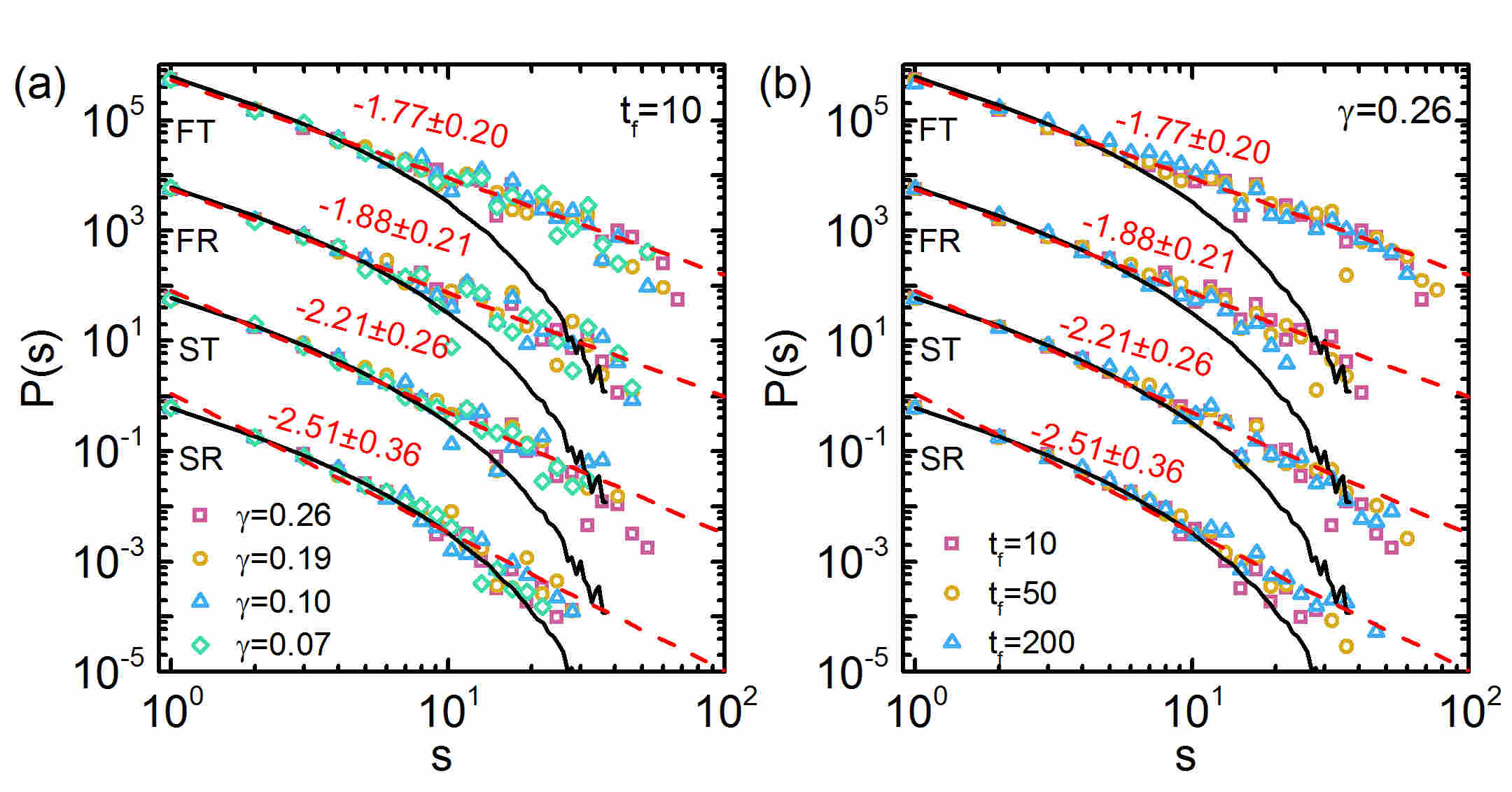}
\caption{
Cluster size distribution $P(s)$ for FT, FR, ST, and SR. (The first three
sets have been shifted vertically.) The solid black
lines are the distributions obtained by randomly picking particles in
the system. The dashed red lines are fits to the data with a power-law
with exponents stated next to these lines. (a) Different strain amplitude
$\gamma$. (b) Different filter times $t_f$.
}
\label{fig2_cluster_size_distr}
\end{figure}

To determine the cluster size distribution $P(s)$ of the four
populations of particles we define two particles to be neighbors
if their Voronoi cells have a common face and use this to construct
connected clusters.  Figure~\ref{fig2_cluster_size_distr}a shows $P(s)$
for the four populations at different strain amplitudes $\gamma$ and one
recognizes that within the accuracy of the data the curves for different
$\gamma$ coincide, i.e.~$P(s)$ is independent of $\gamma$. This somewhat
surprising result is likely related to the fact that for granular systems
the details of the relaxation dynamics are universal and independent of
$\gamma$ due to the particular manner such systems explore their phase
space~\cite{kou_17}.

Also included in Fig.~\ref{fig2_cluster_size_distr}a is a fit to
the data with a power-law, a dependence that has been observed in
previous experiments probing the DH~\cite{keys_07}.  The exponent is small for the FT
clusters and larger for the SR ones (see legend) which shows that the
particles with a slow dynamics belong on average to smaller clusters
that the FT particles. (The average cluster size are 3.92, 3.16, 2.46,
and 2.04 for FT, FR, ST, and SR, respectively.) In the SI we show that
the average volume of the Voronoi cell of slow particles is smaller
than the one of the fast particles, i.e.~the ST and SR clusters are
a bit more densly packed than the ones for FT and FR. This shows that
fast particles prefer to form extended loose clusters which will allow
for cooperative fast motion, similar to the case of glass-forming
liquids~\cite{donati_98}. Also included in the figure are the $P(s)$
obtained if one picks 10\% of particles in a random manner, i.e.~one does
not select fast/slow ones. The so obtained average cluster size is 1.94
and the corresponding $P(s)$, solid black lines, shows at intermediate
and large $s$ the expected exponential decay, i.e.~a $s-$dependence that
is very different from the power-law that we find for the most/least
mobile particles~\cite{footnote2}.

\begin{figure}[tbh]
\centering
\includegraphics[width=0.35\textwidth]{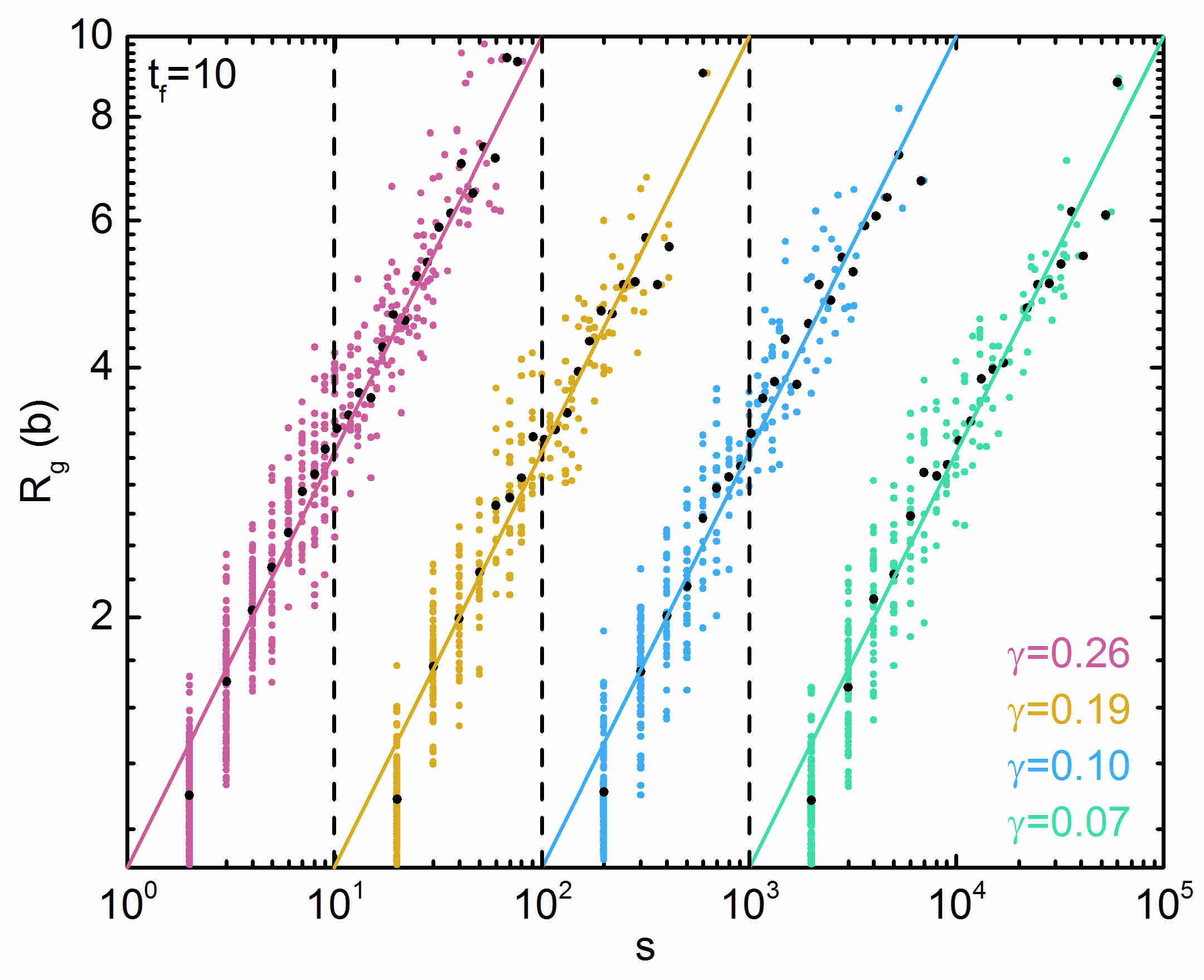}
\caption{
Radius of gyration of the FT clusters vs $s$ for different values of
$\gamma$. Black dots are the average at fixed $s$. The solid lines are
a guide to the eye with slope 0.5.  For the sake of clarity the data for
$\gamma< 0.2$ have been shifted to the right by 10, $10^2$, and $10^3$.
}
\label{fig3_shape_of_clusters}
\end{figure}

In Fig.~\ref{fig2_cluster_size_distr}b we show $P(s)$ for
$\gamma=0.26$ and different values for the filter time $t_f$ and
it is clear that the distribution is also independent of $t_f$.
Therefore Fig.~\ref{fig2_cluster_size_distr} demonstrates that the
DH are independent of the time scale considered, i.e.~$t_f$, and
of the manner the system is driven, $\gamma$.  This is in contrast
to the findings for thermal glass-forming systems in which the DH
are usually found to depend on the time scale considered and also
on the temperature, i.e.~a parameter that is somewhat analogous to
our driving strength $\gamma$. The surprising fact that clusters are
present already at very small $t_f$, also confirmed by the observation
that the non-Gaussian parameter is basically independent of $t$
(Fig.~SI-\ref{figSI_nongaussian}), shows that the DH are {\it not}
related to the $\alpha-$relaxation process, in contrast to the findings
for thermal glass-formers~\cite{donati_98,mazza_07}. Instead, as argued below,
we conjecture that it is the surface roughness of the particles that is
the source of disorder which leads to the DH, i.e.~the same mechanism
that gives rise to the universal relaxation dynamics observed in
Ref.~\cite{kou_17}.

The cluster size distribution $P(s)$ is only related to the number of
particles in a cluster but contains no information about its geometry.
Therefore we determine the radius of gyration $R_g$ of a cluster via\\[-9mm]

\begin{equation}
R_g^2=\frac{1}{s}\sum_{i=1}^s ({\bf r}_i-\overline{{\bf R}})^2 \quad,
\label{eq_rg}
\end{equation}

\begin{figure}[t]
\centering
\includegraphics[width=0.47\textwidth]{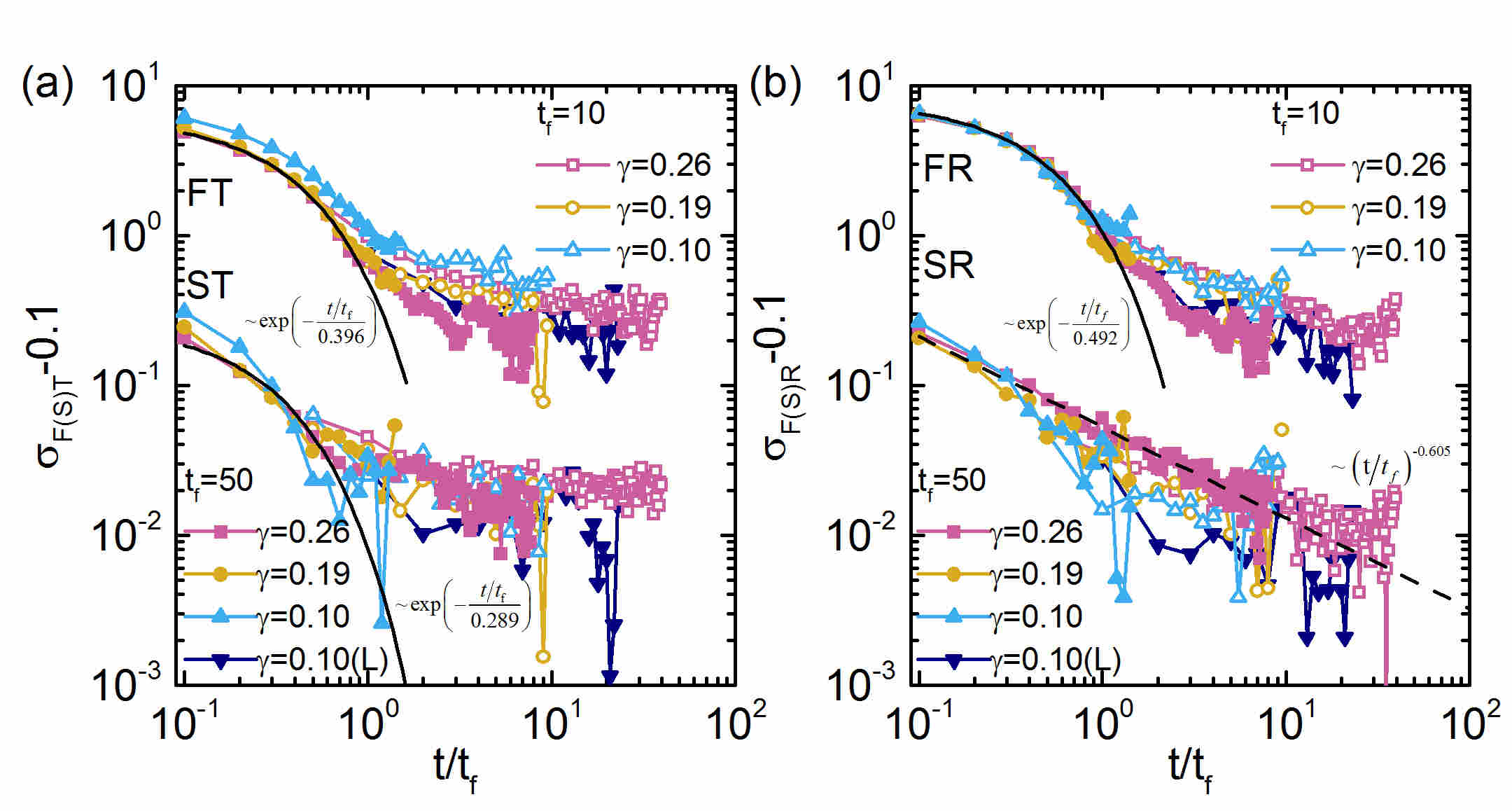}
\caption{
Probability that a particle that was fast/slow at $t=0$ is still fast/slow
at time $t$. Black solid lines are exponential fits to the data at short
times. Panels (a) and (b) are for the TDOF and RDOF, respectively. The
data for SR is fitted with a power-law (black dashed line). The FR and FR data has been 
shifted vertically.
}
\label{fig5_life_time_FS}
\end{figure}

\vspace*{-2mm}
\noindent
where ${\bf r}_i$ is the position of particle $i$ and $\overline{{\bf R}}$
is the center of the cluster. In Fig.~\ref{fig3_shape_of_clusters} we
show how $R_g$ for the FT clusters depends on the cluster size $s$ and
one sees that this dependence is described well by a power-law with an
exponent of 0.5, i.e.~a mass fractal exponent of 2.0~\cite{footnote1}.
This value is independent of the type of particle we consider (FT, FR,
...), see SI, indicating that the geometry of the clusters is independent
of the type of motion considered, in contrast to thermal glass-forming
systems for which one finds that the clusters with FT are more open than
the ones for ST~\cite{chaudhuri_08}. Figure ~\ref{fig3_shape_of_clusters}
also demonstrates that these values are again independent of $\gamma$. If
we select the particles randomly the resulting mass fractal exponent
is around 1.9, i.e.~a value that is very close to the one we find for
the DH clusters. Hence we conclude that the geometry of the DH clusters
is very similar to the one of a random cluster, but that they have an
enhanced probability to be large.

The nature of the particles, FT, FR,..., will change with time and
hence it is of interest to probe how long a particle keeps this
property since this time can be expected to related to the life time
of the clusters. Therefore we define the quantity $\sigma_\alpha(t)$
as the probability that a particle which at time $t=0$ had a property
$\alpha \in \{$FT, ST, FR, SR$\}$ has the same property at time
$t$. In Fig.~\ref{fig5_life_time_FS} we show the $t-$dependence of
$\sigma_\alpha$ for the different  particles. One recognizes that the
curves for the different strain amplitudes fall on top of each other,
i.e.~the persistence $\sigma_\alpha(t)$ is independent of $\gamma$. This
result is surprising since naively one might have expected that a larger
strain would lead to a faster loss of memory because for a given fixed time
$t$ the mean squared displacements of the particles increases quickly
with $\gamma$ (see Fig.~SI-\ref{figSI_msd})~\cite{footnote4}. The figure
also demonstrates that the master curve does not depend on the filter time $t_f$
if one plots the data as a function of the reduced time $t/t_f$. This
independence shows that the details of the relaxation dynamics do not
depend on the time scale considered, i.e.~there is a scale invariance of
the dynamics in the time window we probe, suggesting that the configuration
space explored by the system has a fractal-like nature.

\begin{figure}[tbh]
\centering
\includegraphics[width=0.35\textwidth]{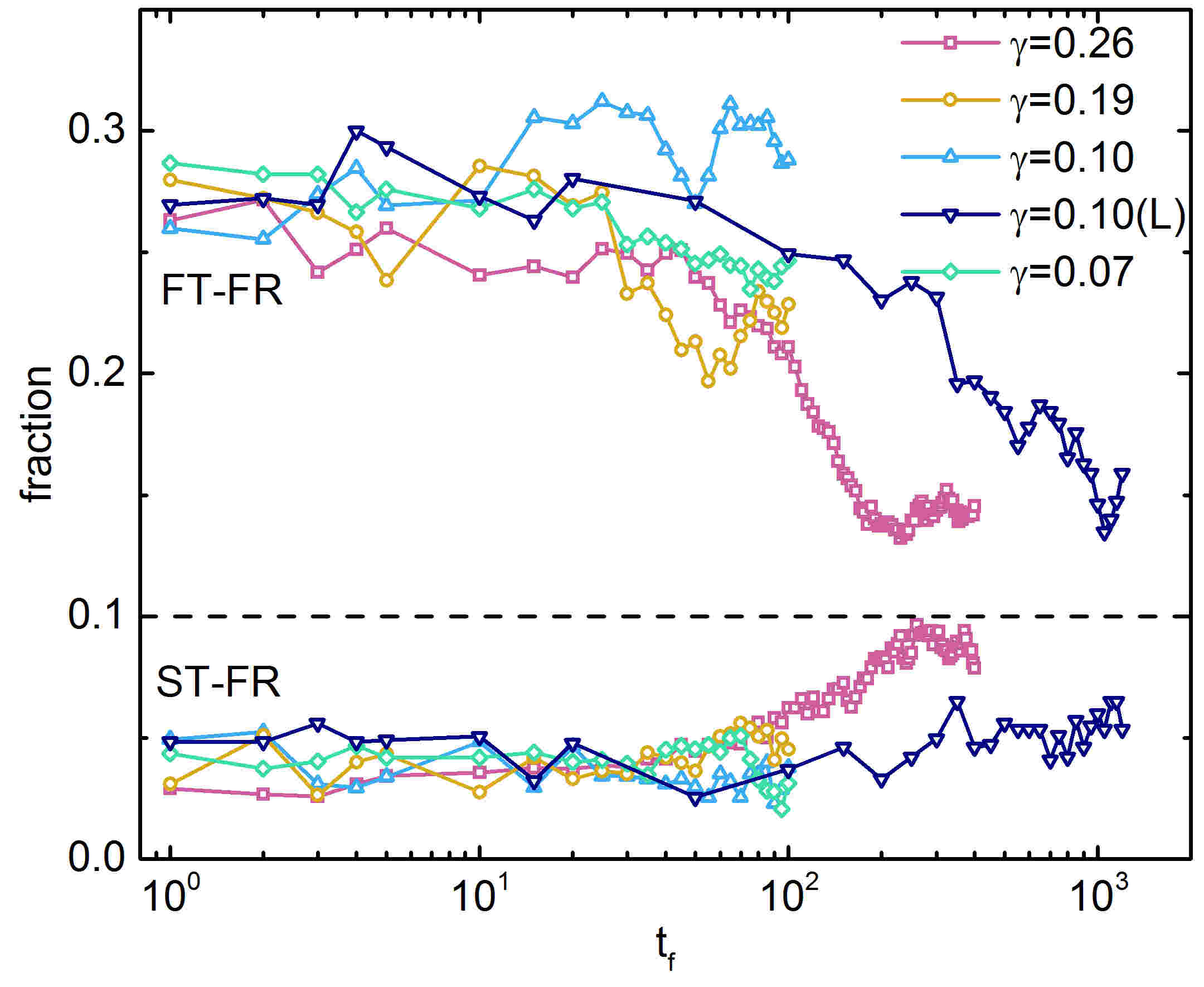}
\caption{
Probability that a particle is undergoing a FT as well as FR (ST and FR) as a
function of the filter time $t_f$. The different symbols correspond to
different strain amplitudes. The dashed line shows the probability at long times.
}
\label{fig4_cluster_overlap}
\end{figure}

Also included in the figure are fits with an exponential function
(black solid lines). These fits describe the data well for $t/t_f \leq
1$ indicating that at short times the particle changes its nature via
a simple stochastic escape process. However, for longer times one observes
a slower decay: For FR and SR we find a power-law with an exponent
around 0.6 whereas for FT and ST the data seems to go to a finite value
given by 0.13 and 0.12, respectively, i.e.~the memory does not vanish
within the time scale of our experiment. The persistence of this memory
regarding the population type of the particle is likely related to the
result discussed in Ref.~\cite{kou_17} where it was found that the TDOF
do show a noticeable memory effect in their motion~\cite{footnote3}.
From panel~(b) we recognize that $\sigma_\alpha(t)$ for the RDOF shows
at intermediate time a power-law with  an exponent around 0.6, i.e.~there
is no plateau, a result that is coherent with the findings that the RDOF
show only a rather weak memory in their dynamics~\cite{kou_phd_18}. From
a physical point of view the absence of a plateau is reasonable since the
cycling motion {\it and the presence of friction} will always lead to a slow
rotation of the particles thus permitting them to change their nature
(FR or SR). 

The persistent rotational motion will induce also a translational
motion of the neighboring particles because of steric effects and
hence we do expect that friction leads to an enhancement of the
rotational-translational coupling~\cite{mohan_02} which in turn
might lead to correlations between the DH of the TDOF with the ones
of the TDOF. To study this correlation we determine the overlap
between the different clusters, i.e.~the probability that a particle
simultaneously belongs to two different populations, e.g.~FT and
FR. In Fig.~\ref{fig4_cluster_overlap} we show this probability for the
combinations FT-FR and ST-FR as a function of $t_f$. (Since the definition
of the population depends on $t_f$, it is clear that the overlap will
depend on $t_f$ as well.) We see that for small and intermediate $t_f$
the overlap for FT-FR is around 0.27, i.e.~by a factor of 2.7 above
the trivial value of 0.1 expected for random clusters, and that this
enhancement is independent of $\gamma$. This implies that there is a
significant probability that a particle which is translating quickly
is also rotating quickly. For $t_f$ larger than around $10^2$ the
overlap starts to decay. From the mean squared angular displacement (see
Fig.~SI-\ref{figSI_msd}) one recognizes that this time scale is related
to the onset of a significant rotational dynamics, i.e.~the particles
have rotated far enough that the nature of their rotational dynamics
has been randomized, thus leading to a decrease of the overlap.  For the
case of the ST-FR we see that the overlap is lowered by a factor of two
with respect to the trivial value, i.e.~slowly translating particles
have a significantly reduced probability to rotate quickly, a result
that is of course very reasonable. This overlap starts to approach the
equilibrium value 0.1 for times that are again on the order of $10^2$
cycles, i.e.~when the particle has rotated by a significant amount
(about 0.5~rad$^2$, see Fig.~SI-\ref{figSI_msd}).  In the SI we show,
Fig.~SI-\ref{figSI_overlap_SR_ST}, that also the pair ST-SR has an
enhanced overlap and the FT-SR has a decreased overlap. None of these
overlaps depend on $\gamma$ if $t_f$ is not too small and only the decay
to the trivial value depends on the driving strength indicating that
the decay is indeed related to the randomization of the RDOF.

The presented results show that our granular system has DH for
the TDOF that are qualitatively similar to the ones found in simple
glass-forming liquids. Having access for the first time to the RDOF
in a 3d experimental system we have probed also the dynamics of the
RDOF and we find that also they do show DH with cluster sizes that
are just a bit smaller than the ones for the TDOF. The cluster size
distribution of all four populations of particles can be described well
by a power-law, thus a distribution that is very different from the
one of random clusters. Remarkably these distributions are independent
of the strain amplitude, indicating that the dynamics is system universal,
in qualitative agreement with earlier findings about the van Hove
function~\cite{kou_17}. The strong correlation and anti-correlation between
different types of clusters shows that in this system the translational
and rotational degrees of freedom are strongly coupled. This coupling
is likely not only caused by the aspect ratio of the particles but also
by the presence of friction, an effect that is absent in molecular
system. 

In Ref.~\cite{kou_17} we argued that 3d granular systems show a relaxation
dynamics that is very different from the one of thermal glass-formers
(e.g.~there is no cage effect).  Although we now find that the DH are
qualitatively similar to the ones of thermal glass-formers, we emphasize
that the DH we observe here occur on the time scale that is significantly
shorter than the $\alpha-$relaxation time, i.e.~the time scale at which
the particles leave their neighborhood.  This surprising observation
shows that the energy landscape of granular materials has a structure
that is very different from the one of thermal glass-formers since it has
a roughness on a length scale that is much smaller than the size of the
particles. It can be expected that it is this particle inherent disorder
that gives rise to the DH, in contrast to the case of thermal systems in
which the variations of the local packing are the cause for the DH. We
expect that it is this roughness which makes the properties of the DH to be
independent of the driving amplitude and the time scale considered.

In summary we conclude that the presence of DH is not a feature that
is unique to thermal glass-formers but instead can be found in other
disordered systems as well and hence is more universal than expected. The
mechanisms leading to these DH are, however, strongly dependent on the
system considered.

\acknowledgements
Some of the preliminary experiments were carried out at BL13W1 beamline
of Shanghai Synchrotron Radiation Facility. The work is supported by the
National Natural Science Foundation of China (No. 11175121, 11675110 and
U1432111), Specialized Research Fund for the Doctoral Program of Higher
Education of China (Grant No. 20110073120073), and ANR-15-CE30-0003-02.

\newpage

{\Large \bf{Supplementary Information}}

\vspace*{10mm}
{\bf Experimental details}\\

The prolate ellipsoid particles are made of polyvinyl chloride and
the 4100 particles, with a total weight of 8.3kg, were poured into the
shear cell. A plate that is constrained to move only in the vertical
direction was placed on top of the particles. This top plate, with
weight of 16kg, provided an extra constant normal pressure of 2.3kPa
on the top surface of the packing and thus makes that the pressure
gradient in the vertical direction is significantly reduced. During an
oscillatory shear cycle, the shear cell will deform to a designed shear strain,
then in the opposite direction to the same shear strain and finally return to the
origin position. The strain rate for the preparation of the sample
and the subsequent measurements is around $1.7\cdot 10^{-2}$s$^{-1}$
so that the inertial number $I=2\dot{\gamma}b/\sqrt{P/\rho}$ is
about $1.4\cdot 10 ^{-4}$, i.e.~the experiment can be considered as
quasi-static~\cite{gdr_midi_04}. (Here $P$ is the pressure and $\rho$ the
mass density.) Before starting the computational tomography (CT) scans
we drove the system into a steady state by making hundreds of shear cycles.
The number of cycles for the preparation and the subsequent measurements are
given in Supplementary Table~1.

Using a medical CT scanner (SOMATOM Perspective, Siemens, Germany), we
obtained the complete three-dimensional structural information of the
packing with a spatial resolution of 0.6mm. Following similar imaging
processing steps as in previous studies, Refs.~\cite{kou_17,xia_14},
we determined the position and orientation  of each ellipsoid by a
marker-based watershed image segmentation technique. We estimated the
precision of the position ${\bf r}_i$ and the orientational vector
$\boldsymbol{\mbox{e}}_i$ of a particle $i$ by making two consecutive
tomography scans of the same static packing and comparing the difference,
which gave $5.3\cdot 10^{-3}b$ and $8.4 \cdot 10^{-3}$rad, respectively.
To avoid the influence of boundary effects, we excluded all particles
that have a distance less than $5b$ from the cell boundary so that
we only considered about 1300 central particles for analysis. See
Ref.~\cite{kou_17} for more details.

To determine the dynamic properties of the system we took a tomography
scan after each cycle for the first 10 cycles and then a scan every
5 cycles. For $\gamma = 0.10$ we made a second experiment in which we
scanned only every 50 cycles allowing thus to reach larger times. The
so obtained results are labeled as ``0.10(L)''.

From the positions and orientations of the particles we calculated
their translational displacement as ${\bf d}_j(t)={\bf r}_j(t)-{\bf
r}_j(0)$ while the rotational displacement was obtained
from the time integral of the angular increment of each cycle,
$\boldsymbol{\theta}_j(t) = \int_0^t \boldsymbol{\omega}_j(t') dt'$,
where the modulus and direction of $\boldsymbol{\omega}_j(t)$ are given by
$\cos^{-1}[\boldsymbol{\mbox{e}}_j(t)\cdot \boldsymbol{\mbox{e}}_j(t+1)]$
and the vector $\boldsymbol{\mbox{e}}_j(t)\times
\boldsymbol{\mbox{e}}_j(t+1)$, respectively.

\begin{table}[t]
{\centering
 \begin{tabular}{|p{25mm}|p{25mm}|p{25mm}|}
\hline
Shear strain \newline (y-direction) & 
Number of cycles to prepare initial state &
Number of cycles for measurements \\
\hline
0.26 & 300 & 615 \\
0.19 & 550 & 125 \\
0.10 & 1500 & 125\\
0.10 (L) & 1500 & 1850\\
0.07 & 2400 & 125\\
\hline
 \end{tabular}
\caption{
Experimental protocol used to prepare the system
and to make the measurements of its properties.}
\label{table-shear}
}
\end{table}

\vspace*{10mm}
{\bf Static quantities}\\[1mm]

\begin{figure}[tbh]
\includegraphics[width=0.3\textwidth]{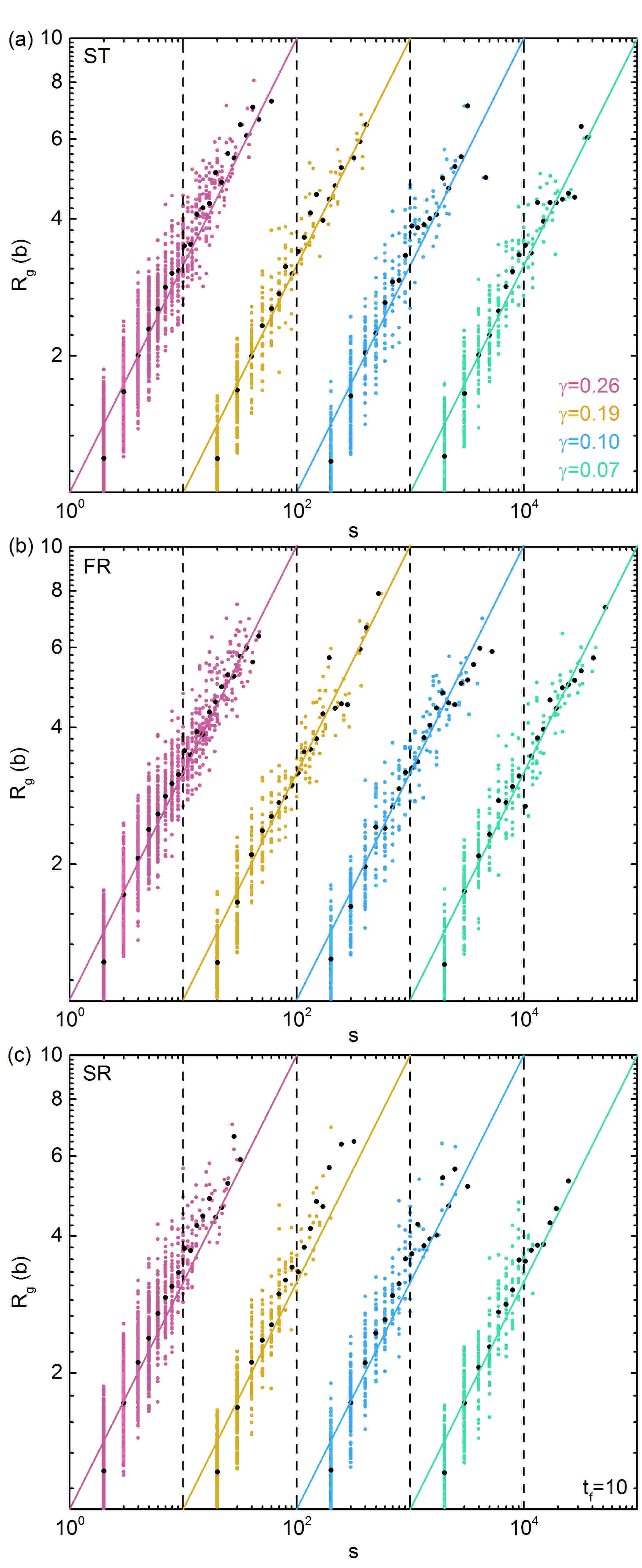}
\caption{
}
Radius of gyration of the clusters as a function of cluster size $s$ for different
values of $\gamma$. (a): ST, (b) FR, and (c) SR. The solid lines are a
guide to the eye with slope 0.5. Black dots are the average at fixed $s$. 
For the sake of clarity the data for
$\gamma< 0.2$ have been shifted to the right by 10, $10^2$, and $10^3$.
\label{figSI_Rg_ST_FR_SR}
\end{figure}

In Fig.~SI-\ref{figSI_Rg_ST_FR_SR} we show the relationship between radius
of gyration $R_g$ and cluster size $s$ for the ST (panel a), FR (panel
b) and SR cluster (panel c), which gives an $s-$dependence that is
very similar to the one of the FT cluster as presented in Fig.~3 of the
main text, i.e.~a power-law with an exponent close to 2.0. Thus we can
conclude that the geometric shape of the clusters is independent of the
nature of the particles considered.

\begin{figure}[tbh]
\includegraphics[width=0.47\textwidth]{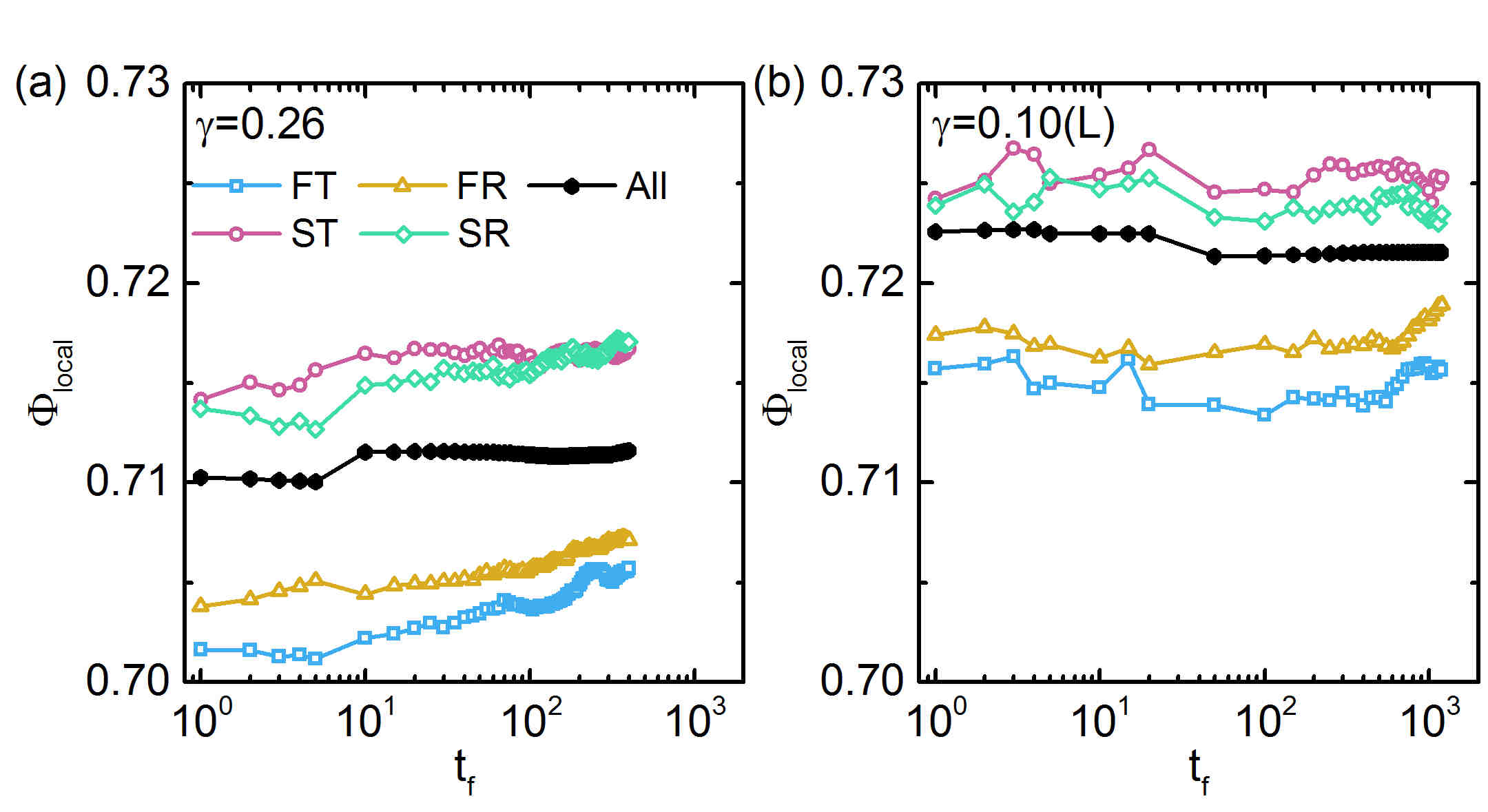}
\caption{
}
The average local volume fraction $\Phi_{\rm local}$
for particles that are FT, FR, ST, and ST as a function of the filter
time $t_f$. Also included is the sample averaged local volume fraction
(curve labeled ``All''). Panels (a) and (b) correspond to $\gamma=0.26$
and $\gamma=0.1$(L), respectively.  
\label{figSI_phi_vs_tf}
\end{figure}

In Fig.~SI-\ref{figSI_phi_vs_tf} we show the $t_f$-dependence of the
local volume fraction $\Phi_{\rm local}$ for different types of
particles. $\Phi_{\rm local}$ is obtained by calculating for each
particle in the population considered $\Phi_{{\rm local},i}=V_{\rm
part}/V_{{\rm Vor},i}$, where $V_{\rm part}$ is the volume of
a particle and $V_{{\rm Vor},i}$ is the volume of the Voronoi
cell of particle $i$, and then taking the average of this ratio over the particles.
The graphs show that for $\gamma=0.26$, panel (a), particles with FT and
FR have a $\Phi_{\rm local}$ that is about 1\% smaller than the average,
whereas the ST and SR particles have a $\Phi_{\rm local}$ that is about
1\% higher. Qualitatively the same
result is obtained for the strain amplitude $\gamma=0.1$(L), panel (b),
but now the dependence on the type of particle is somewhat weaker. We also note
that in both cases the $t_f$-dependence is relatively weak, thus showing
that, for fixed $\gamma$, the filter time does not strongly influence
the local packing density of the selected particles.

\begin{figure}[tbh] 
\includegraphics[width=0.4\textwidth]{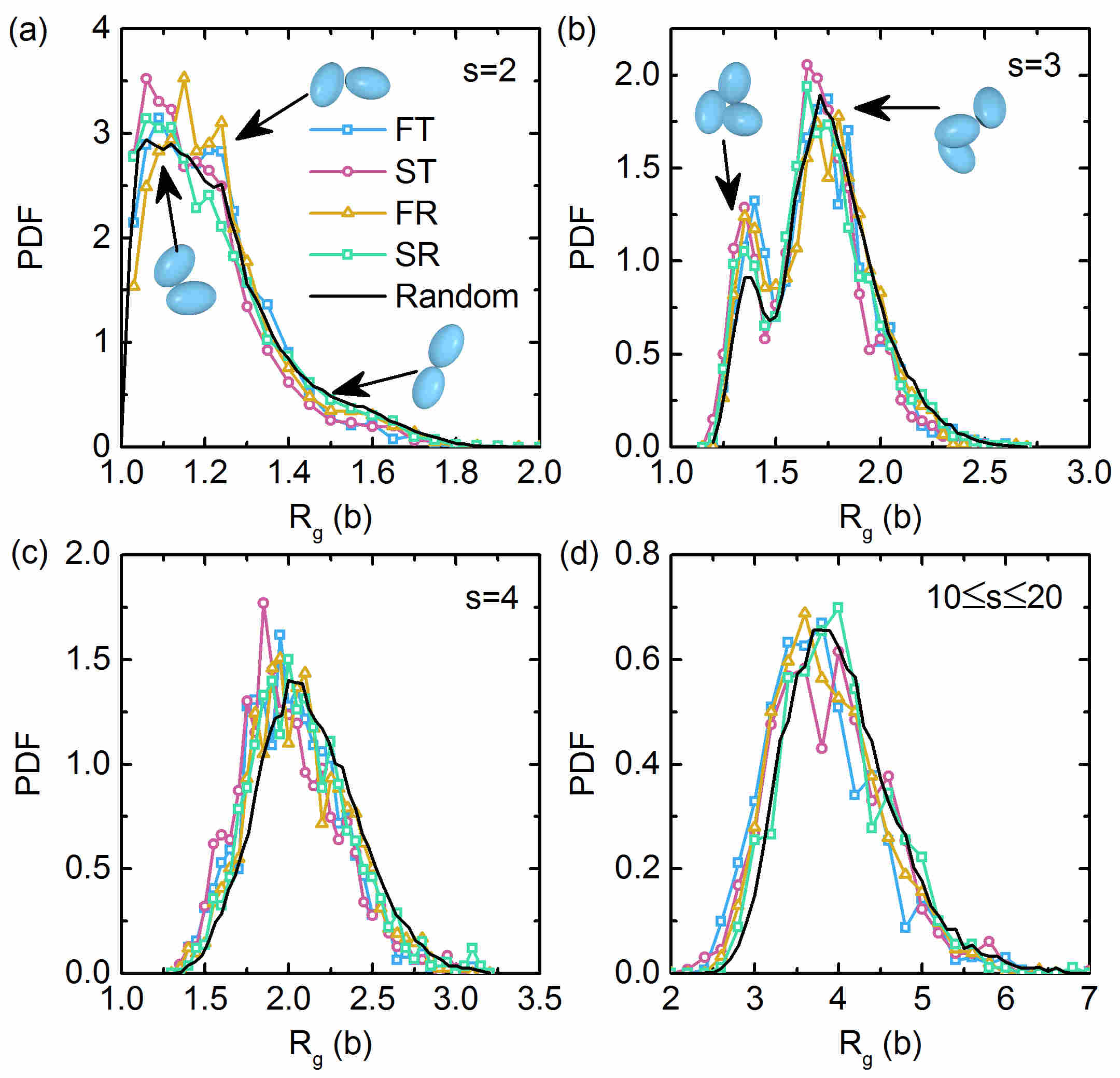}
\caption{
Probability distribution function of $R_g$ of FT, ST, FR, SR, and random
clusters, for different cluster sizes $s$: (a): $s=2$; (b) $s=3$; (c) $s=4$
(d) $10\leq s\leq 20$.
}
\label{figSI_rg_cluster}
\end{figure}

\begin{figure}[tbh]
\includegraphics[width=0.45\textwidth]{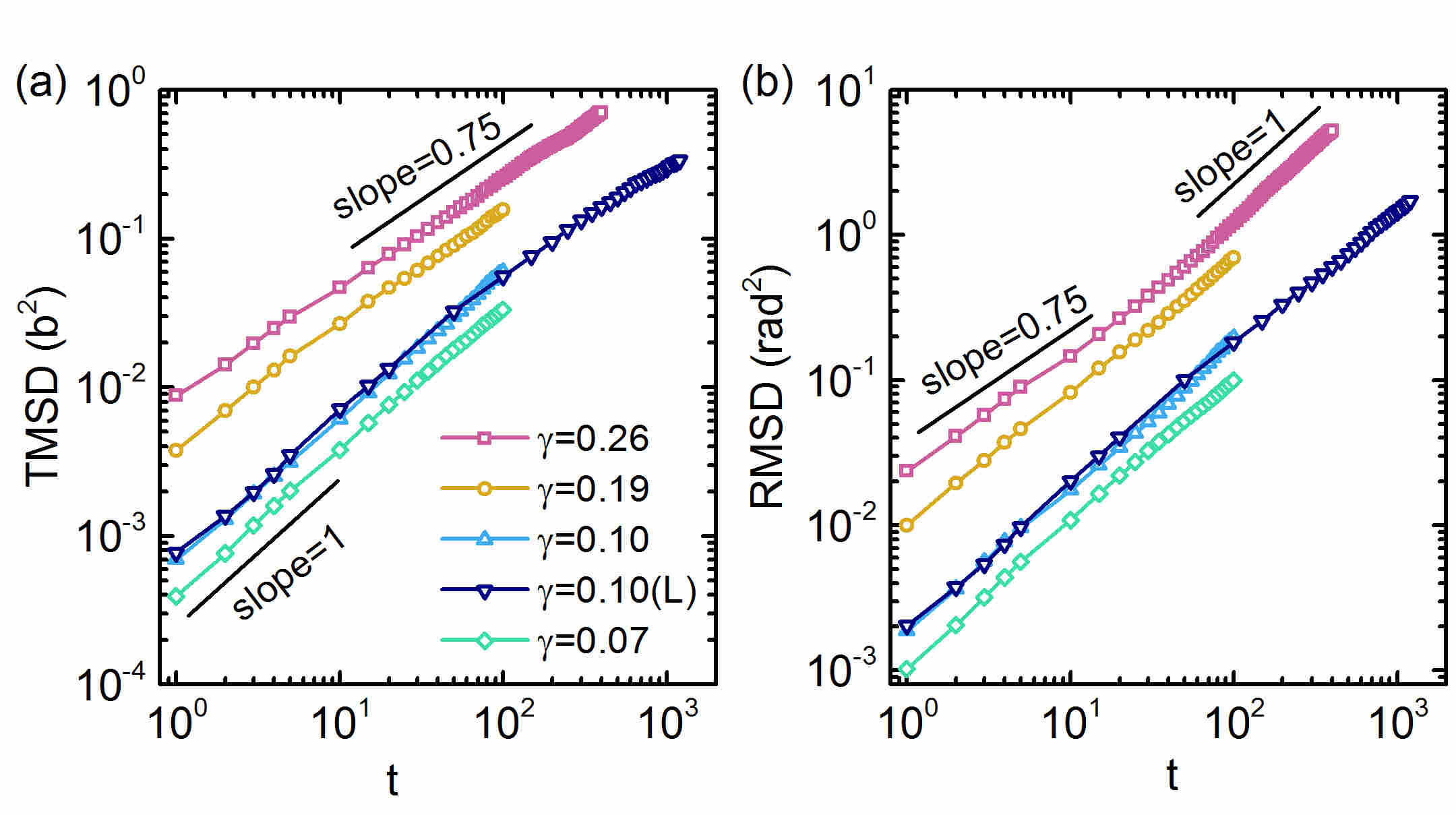}
\caption{
Time dependence of the TMSD and RMSD for different strain amplitudes $\gamma$.
Adapted from Ref.~\cite{kou_17}.
}
\label{figSI_msd}
\end{figure}

In Fig.~3 of the main text we have shown the radius of gyration of
the FT clusters and in Fig.~SI-\ref{figSI_Rg_ST_FR_SR} the $R_g$ for the
other types of clusters. In Fig.~SI-\ref{figSI_rg_cluster} we show the
corresponding distributions for selected values of the cluster size $s$.
For very small clusters, $s=2$ in panel (a), we see that the distribution
has a broad peak with a shape that is directly related to the manner
the two particles touch each other (see cartoons of the configurations in
the figure). No significant dependence on the cluster type is seen, but
a comparison with the distribution of the random clusters (black line)
shows that the latter has a slightly more pronounced tail at large $R_g$,
i.e.~configurations in which the particles touch each other at their
tips. That the DH clusters have a less pronounced tail is likely related
to the fact that in that geometry (see cartoon) the neighboring particles
barely touch and hence friction is not able to couple their motion.

For the case $s=3$, panel (b), we find a double peak structure in the
distribution. A closer inspection of the configurations reveals that the peak at
smaller distance corresponds to an arrangement in which all three
particles touch other two whereas the peak at larger distance
to the case in which only two particles touch a central one (see
cartoons in the figure). Comparing the distributions for the different
cluster types clearly shows that the random cluster has a significantly
smaller probability of having three particles touching each other, a result
that is very reasonable from the combinatorial point of view and the
fact that a tight packing increases the dynamic coupling between the
particles. Thus we can conclude that for this cluster size the non-random
clusters are more compact than the random ones.

For clusters sizes that are equal or larger than $s=4$ the shape of
the distribution changes significantly in that it becomes Gaussian
like (panels (c) and (d)). We see that also in these cases the random
distribution peaks at slightly larger values of $R_g$, i.e.~the DH
clusters are slightly more compact than the random ones. It is also
interesting to note that within the accuracy of our data the distribution
for the four types of clusters is independent of the mobility of the
particles, a result that is consistent with the ones presented in Fig.~3
of the main text and Fig.~SI-\ref{figSI_Rg_ST_FR_SR}.

\vspace*{10mm}
{\bf Dynamical quantities}\\[1mm]
To comprehend how far the particles translate and rotate within the
time scale of our experiment we show in Fig.~SI-\ref{figSI_msd}~(a)
and (b) the mean squared displacement of the particles for the TDOF
and RDOF, respectively. The data is the same as the one shown in
Ref.~\cite{kou_17}.  Note that the RDOF reach the diffusive limit
significantly earlier than the TDOF, i.e.~the former are faster than
the latter. We also recognize a strong $\gamma-$dependence of the data,
making the $\gamma-${\it independence} of the results shown in the main
text rather surprising.

\begin{figure}[bh]
\includegraphics[width=0.5\textwidth]{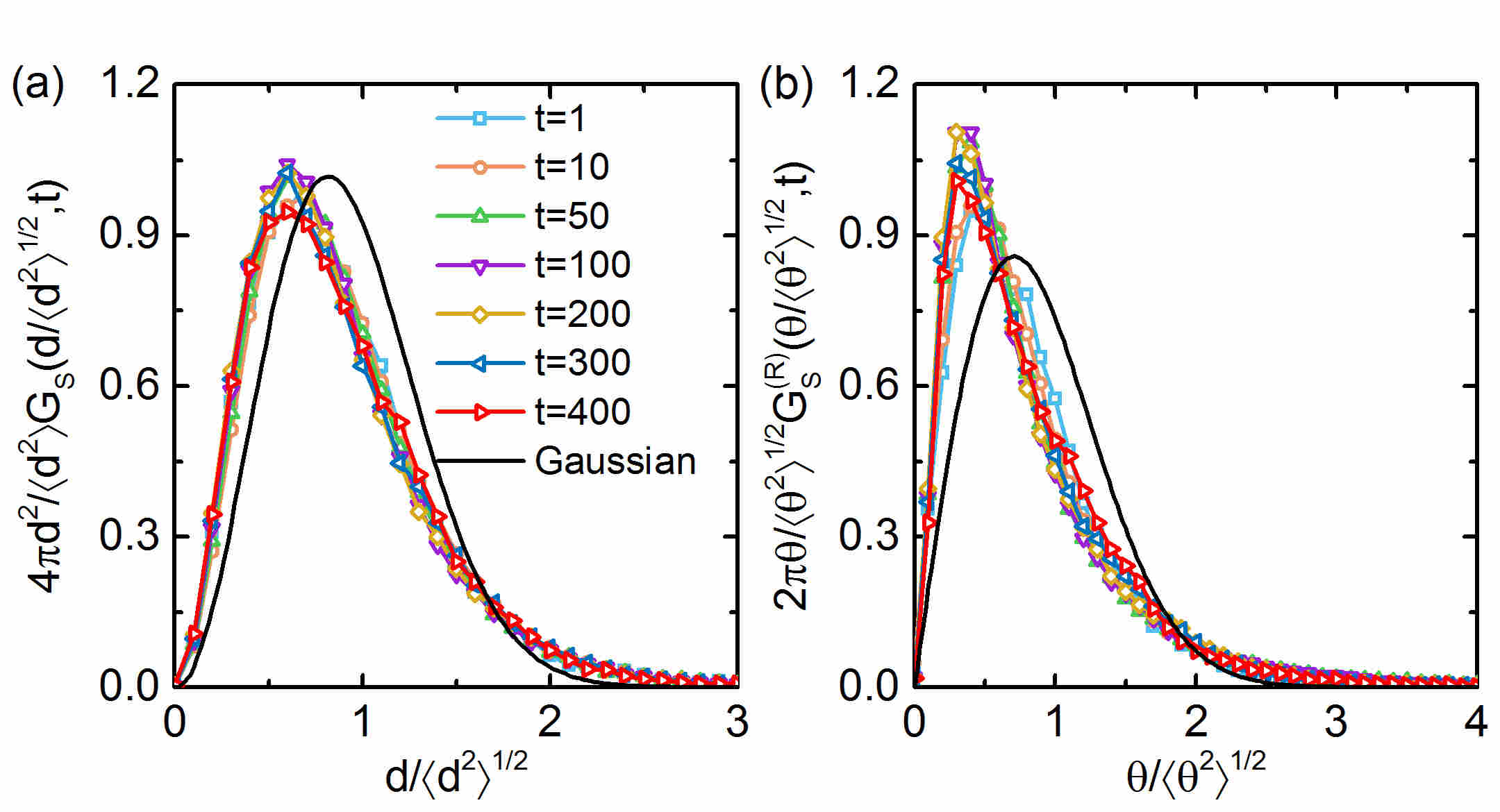}
\caption{
Self part of the van Hove function for the TDOF and RDOF (panel (a)
and (b), respectively) for $\gamma=0.26$. The different symbols are for
different times $t$ and the distance shown on the abscissa is normalized
by the corresponding square root of the mean squared displacement. The
black solid line corresponds to a Gaussian distribution.
}
\label{figSI_gs}
\end{figure}

In Fig.~SI-\ref{figSI_gs}~(a) and (b) we show the self-part of the van
Hove function for the TDOF, $G_s(r,t)$ (see Ref.~\cite{kou_17} for
its definition), and RDOF, $G_s^{(R)}(\theta,t)$, respectively. Here
$G_s^{(R)}(\theta,t)$ is defined as $G_s^{(R)}(\theta,t)=N^{-1}
\sum_{j=1}^N \langle \delta(\theta - |\theta_j(t)-\theta_j(0)|)\rangle$,
where $N$ is the number of particles and $\delta$ is the Dirac
$\delta-$function. The different symbols correspond to different
times $t$.  Note that the abscissa is scaled by the square root of the
mean squared displacement at the selected times which makes that the
distribution functions fall onto a master curve, a non-trivial result
that is discussed in detail in Ref.~\cite{kou_17}. Also included in the
figures are Gaussians with identical means as the real data (black solid
lines). The Gaussians cross the data at around $\xi=$1.65(T) and 1.75(R).
Thus it is reasonable to define those particles as ``fast'' that
have displacements larger than $\xi$.  If one integrates the area under
the data from $\xi$ to infinity one gets values between 7.3 and 9.6\%
i.e.~values that are close to the one used in our definition for fast
particles, i.e.~the top 10\% of the distribution. (We mention that we have
also repeated the analysis of the cluster size etc.~by defining the
fast/slow particles as the top/bottom 7\% of the distribution and found
no significant difference to the results presented in the main text.)

\begin{figure}[tbh] 
{\centering
\includegraphics[width=0.4\textwidth]{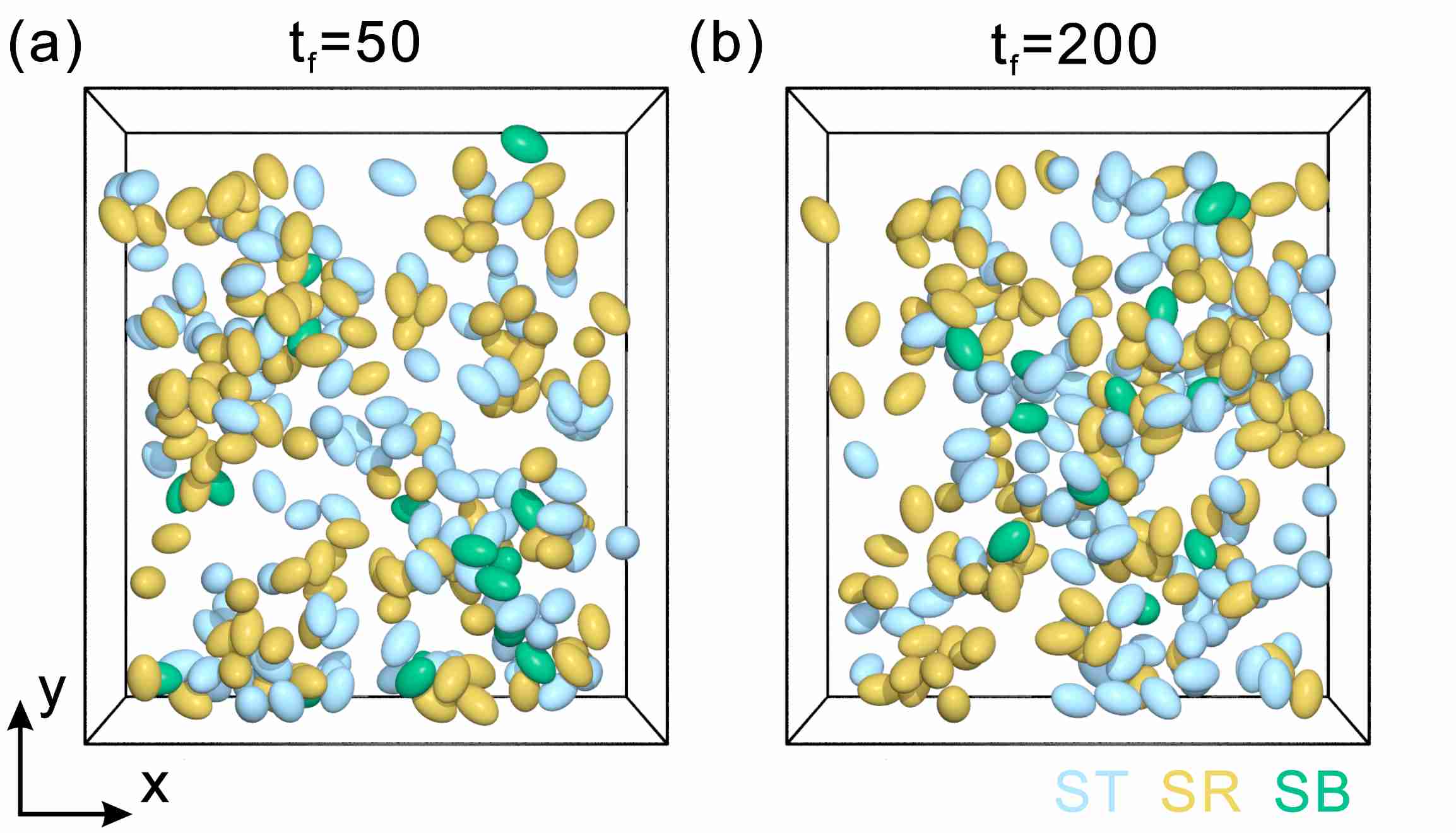}
}
\caption{
Snapshots showing the slow translating (ST, violet) and slow rotating (SR,
orange) particles for $\gamma=0.10$(L) at $t =1000$. Also shown are the particles
that are ST as well as SR (SB, green). Note that both type of particles
are forming clusters and that these clusters overlap significantly. Panels
(a) and (b) are for the filter time $t_f = 50$ and $t_f = 200$, respectively.
}
\label{figSI_ST_SR_different_tf}
\end{figure} 

In Fig.~SI-\ref{figSI_ST_SR_different_tf} we show a typical snapshot of
the ST and SR particles.  (The parameters are the same as the ones
of Fig.~\ref{fig1_FT_FR_snapshots} of the main text.) We see that,
as discussed in the main text, also the slowly translating/rotating
particles form clusters, i.e.~DH, and that these two types of clusters
do indeed overlap significantly. This is demonstrated quantitatively in
Fig.~SI-\ref{figSI_overlap_SR_ST} where we show the $t_f$ dependence of the
overlap between ST and SR clusters as well as FT and SR clusters. The
former overlap is about 1.5 higher than the trivial values 0.1 whereas
the latter is about 50\% less than the trivial value. Hence we can
conclude that these clusters are clearly correlated/anticorrelated,
in agreement with the results from Fig.~\ref{fig4_cluster_overlap}.

\begin{figure}[tbh] 
{\centering
\includegraphics[width=0.4\textwidth]{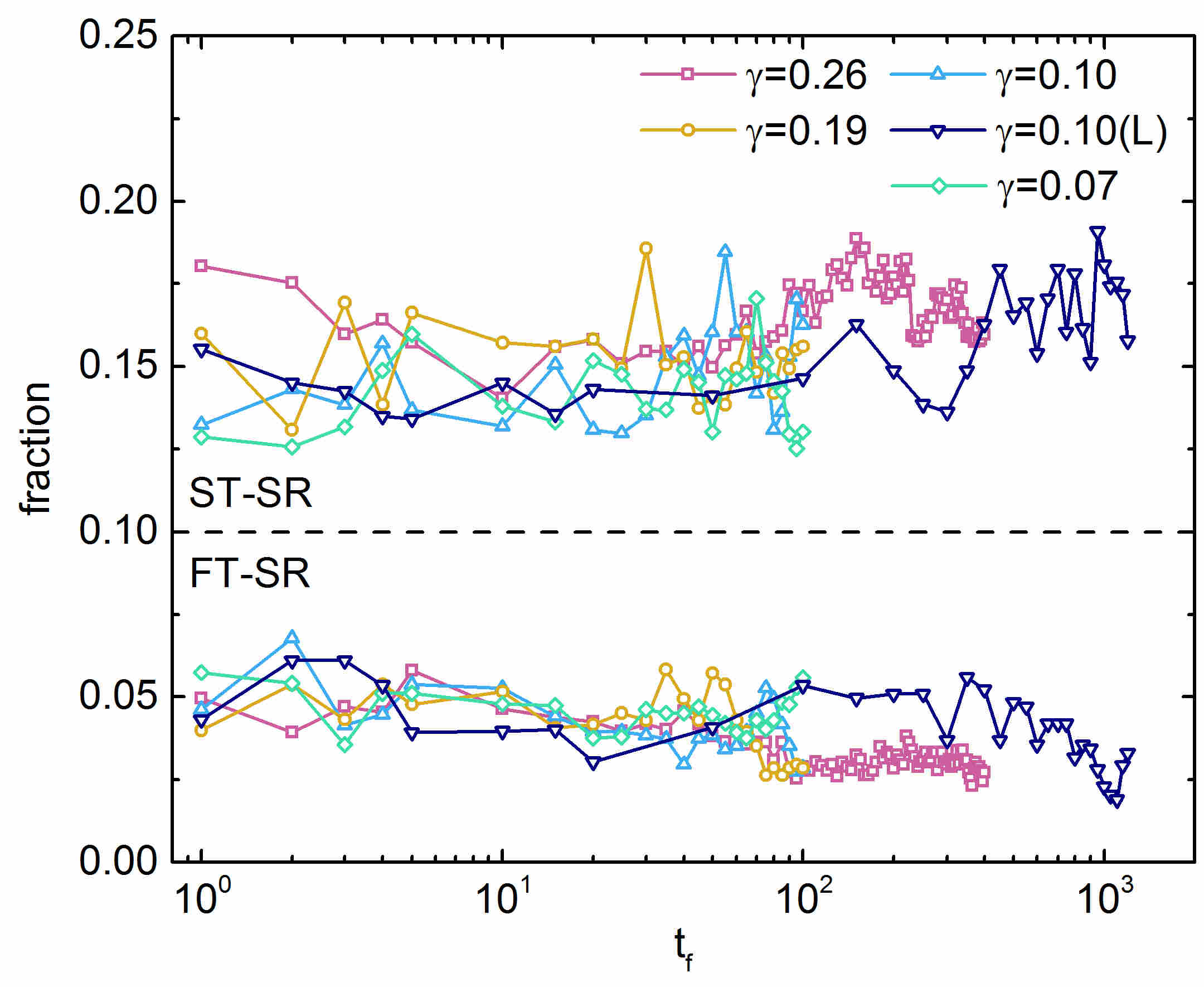}
}
\caption{
Probability that a particle is making a ST as well as a SR (FT and SR)
as a function of the filter time $t_f$. The different lines correspond
to different strain amplitudes. The dashed horizontal line shows the random
value, i.e.~the case that clusters are uncorrelated.
}
\label{figSI_overlap_SR_ST}
\end{figure}

Finally we show in Fig.~SI-\ref{figSI_nongaussian} the time dependence of
the non-Gaussian parameter of the van Hove function for the TDOF, i.e.

\begin{equation}
\alpha_2^{(T)}(t) = \frac{3 \langle r^4(t)\rangle}{5\langle r^2(t)\rangle^2} -1 \quad ,
\label{eq_alpha2}
\end{equation}

\noindent
with an analogous definition for the RDOF. (For the RDOF the
factor 3/5 has to be replaced by 1/2 since the rotational motion is
two-dimensional.) We see, panel (a), that for the TDOF $\alpha_2^{(T)}(t)$
differs significantly from zero already at $t=1$, i.e. the distribution
function of the displacements is not a Gaussian, in agreement with the
results from Ref.~\cite{kou_17}. For larger times $\alpha_2^{(T)}(t)$
increases slightly, but the $t-$dependence remains weak. We also note that
this quantity is independent of the strain amplitude, also this in qualitative
agreement with the results presented in Ref.~\cite{kou_17}. For the RDOF,
we find that $\alpha_2^{(R})$ is qualitatively similar to $\alpha_2^{(T)}$
except that the data is more noisy, see panel (b).

\begin{figure}[tbh]
\centering
\includegraphics[width=0.45\textwidth]{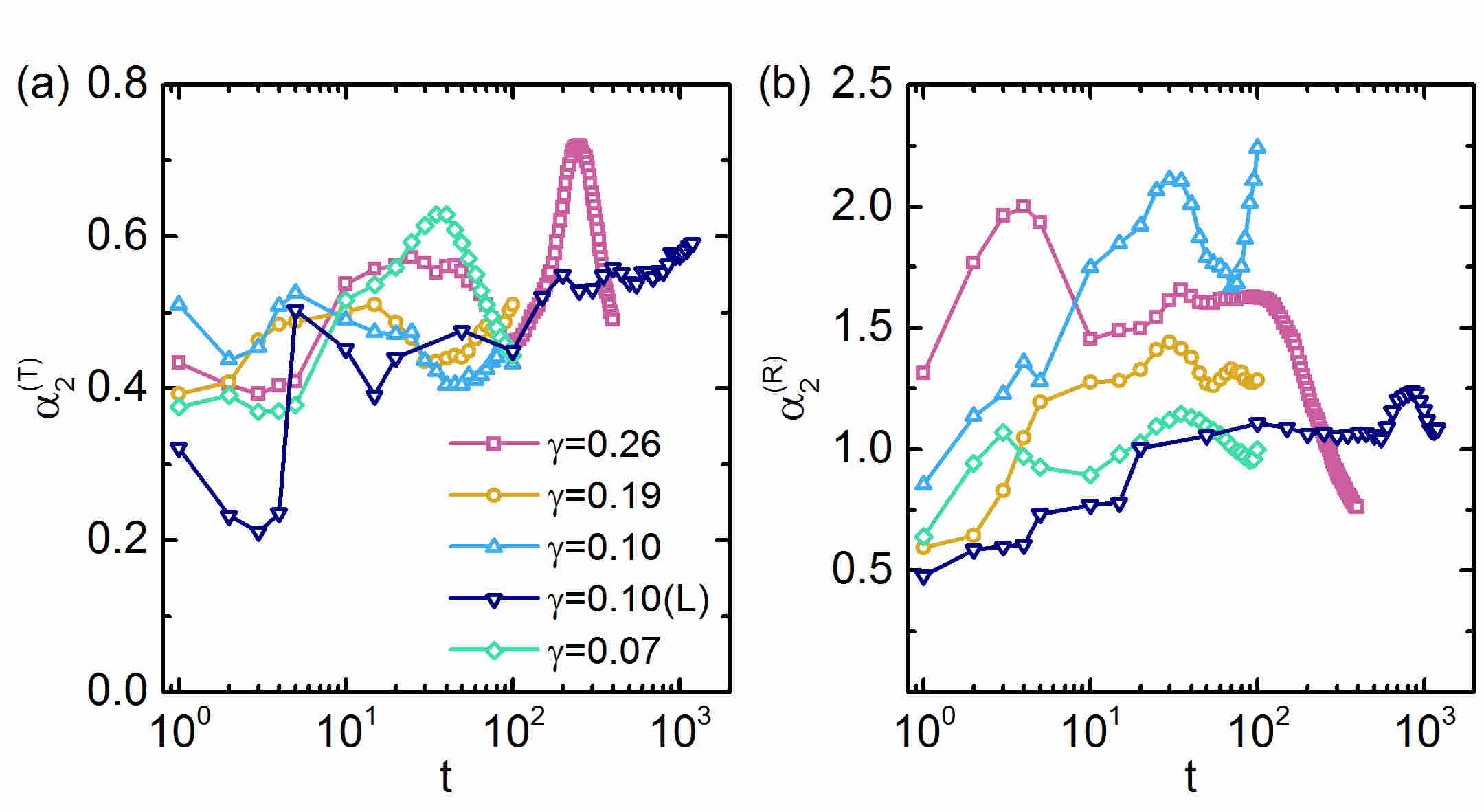}
\caption{
}
Time dependence of the non-Gaussian parameter $\alpha_2(t)$ for the TDOF,
(a), and the RDOF, (b), for different strain amplitudes $\gamma$.
\label{figSI_nongaussian}
\end{figure}


\begin{thebibliography}{10}
\bibitem{footnote2}
We have also checked that the presented results are independent of
the shape of the box considered for the analysis, which show that our
results on the cluster size are robust and independent of the details
on how they have been obtained.
\bibitem{footnote1}
Note that in order to determine this exponent we have neglected the
clusters of size 2 and 3 since their size is mainly given by local
packing geometries (see Fig.~SI-\ref{figSI_rg_cluster}).
\bibitem{footnote4}
Note that this independence is qualitatively different from the one
discussed in Ref.~\cite{kou_17} where it was found that the van Hove
function is independent of $\gamma$ {\it if} one rescales the displacement
by a {\it $\gamma-$dependent} factor.
\bibitem{footnote3}
For times that are well
beyond the $\alpha-$relaxation time it is clear that $\sigma_\alpha(t)$
has to go to the trivial values 0.1 since the system will be completely
randomized, but these time scales are not accessible in our experiments.
\end{thebibliography}
\end{document}